\documentclass[11pt,twocolumn]{article}

\usepackage[margin=1in]{geometry}
\usepackage[T1]{fontenc}
\usepackage{lmodern}
\usepackage{microtype}
\usepackage{authblk}
\usepackage{booktabs}
\usepackage{tabularx}
\usepackage{graphicx}
\usepackage{float}
\usepackage{dblfloatfix}
\usepackage{cuted}
\usepackage{capt-of}
\usepackage{placeins}
\usepackage{flushend}
\usepackage{xcolor}
\usepackage{xurl}
\usepackage{listings}
\usepackage[backend=biber,style=authoryear,sorting=nyt,maxcitenames=2,mincitenames=1,uniquelist=false,natbib=true]{biblatex}
\definecolor{paperlinkblue}{HTML}{2F6FB0}
\usepackage[colorlinks=true,linkcolor=paperlinkblue,citecolor=paperlinkblue,urlcolor=paperlinkblue]{hyperref}
\usepackage{orcidlink}
\usepackage{fontawesome5}

\definecolor{githubblue}{HTML}{0969DA}
\definecolor{promptblue}{HTML}{E6F4FA}
\DeclareDelimFormat{nameyeardelim}{\addspace}
\DeclareDelimFormat{finalnamedelim}{\addspace\&\space}
\renewbibmacro*{volume+number+eid}{%
  \printfield{volume}%
  \iffieldundef{number}
    {}
    {\setunit*{\addspace}\printtext[parens]{\printfield{number}}}%
  \setunit{\bibeidpunct}%
  \printfield{eid}}
\AtBeginBibliography{\scriptsize\linespread{0.94}\selectfont}

\hypersetup{pdftitle={Minimal Local Simulation Foundations for LLM- and VLM-Driven Agents in 2D and 3D Environments},pdfauthor={Ryuki Hyodo}}

\newcommand{\agentfoundrytwourl}{https://github.com/ryukih/SD-AgentFoundry-2D}
\newcommand{\agentfoundrythreeurl}{https://github.com/ryukih/SD-AgentFoundry-3D}
\newcommand{\agentfoundryname}{\textit{SD-AgentFoundry}}
\newcommand{\agentfoundrytwotext}{\textit{SD-AgentFoundry-2D}}
\newcommand{\agentfoundrythreetext}{\textit{SD-AgentFoundry-3D}}
\newcommand{\agentfoundrytwoname}{\href{\agentfoundrytwourl}{\agentfoundrytwotext{}}}
\newcommand{\agentfoundrythreename}{\href{\agentfoundrythreeurl}{\agentfoundrythreetext{}}}
\newcommand{\agentfoundrytwoplainname}{\href{\agentfoundrytwourl}{SD-AgentFoundry-2D}}
\newcommand{\agentfoundrythreeplainname}{\href{\agentfoundrythreeurl}{SD-AgentFoundry-3D}}

\lstdefinestyle{promptstyle}{
  basicstyle=\ttfamily\footnotesize,
  breaklines=true,
  breakatwhitespace=false,
  backgroundcolor=\color{promptblue},
  columns=fullflexible,
  frame=single,
  keepspaces=true,
  showstringspaces=false,
  aboveskip=0.8\baselineskip,
  belowskip=0.8\baselineskip
}

\title{\bfseries Minimal Local Simulation Foundations for LLM- and VLM-Driven Agents in 2D and 3D Environments}

\author[1,2,3,4]{\textbf{Ryuki Hyodo}\,\orcidlink{0000-0003-4590-0988}}
\affil[1]{SpaceData Inc., 1-17-1 Toranomon, Minato-ku, Tokyo 105-6490, Japan}
\affil[2]{Graduate School of Artificial Intelligence and Science, Rikkyo University, 3-34-1 Nishi-Ikebukuro, Toshima-ku, Tokyo 171-8501, Japan}
\affil[3]{Earth-Life Science Institute, Institute of Science Tokyo, 2-12-1 Ookayama, Meguro-ku, Tokyo, 152-8550, Japan}
\affil[4]{Universit\'{e} Paris Cit\'{e}, Institut de Physique du Globe de Paris, CNRS, F-75005 Paris, France}
\affil[ ]{Email: \href{mailto:ryuki.hyodo@spacedata.co.jp}{\texttt{ryuki.hyodo@spacedata.co.jp}}; \href{mailto:hyodo@elsi.jp}{\texttt{hyodo@elsi.jp}}}
\date{}

\begin{document}

\raggedbottom

\makeatletter
\twocolumn[
\begin{@twocolumnfalse}
\maketitle

\begin{abstract}
Large language models (LLMs) and vision-language models (VLMs) are expanding the range of behaviors that can be represented in agent-based simulations, but many contemporary platforms are difficult to study, modify, or run on ordinary computers. We present two intentionally minimal simulation foundations for education and rapid prototyping. \agentfoundrytwoname{} provides a two-dimensional multi-agent environment in which locally hosted LLM agents move, communicate, respond to place occupancy, and encounter spatially localized fire events. \agentfoundrythreename{} provides a three-dimensional digital-twin environment in which a locally hosted VLM receives first-person images and produces natural-language movement instructions. Both codebases are designed to run locally on macOS, Windows, and Linux and are deliberately left open to modification rather than developed as finished applications. Together, they offer accessible starting points for learning about generative social simulation and for building domain-specific extensions.
\end{abstract}

\begin{center}
\begin{minipage}{0.88\textwidth}
\small
\noindent\faGithub\ \textbf{\agentfoundrytwoplainname{}:} \href{\agentfoundrytwourl}{\textcolor{githubblue}{\texttt{https://github.com/ryukih/SD-AgentFoundry-2D}}}

\smallskip

\noindent\faGithub\ \textbf{\agentfoundrythreeplainname{}:} \href{\agentfoundrythreeurl}{\textcolor{githubblue}{\texttt{https://github.com/ryukih/SD-AgentFoundry-3D}}}
\end{minipage}
\end{center}

\vspace{1em}
\end{@twocolumnfalse}
]
\makeatother

\section{Introduction}
\label{sec:introduction}

Agent-based simulation has long provided a bottom-up way to study how individual decisions and interactions can produce collective patterns \citep{bonabeau2002abm}. Recent progress in foundation models has widened this approach: instead of specifying every behavior as a fixed rule, a simulator can ask an LLM to interpret a situation, use natural-language memory, communicate with other agents, and choose an action \citep{guo2024masurvey}. Social Simulacra demonstrated LLM-supported prototyping of populated social-computing systems \citep{park2022social}, while Generative Agents combined memory, reflection, and planning in an interactive two-dimensional town \citep{park2023generative}. Concordia generalized language-mediated agent-based modeling through components that connect LLM calls, associative memory, and an environment-controlling game master \citep{vezhnevets2023concordia}. More recent systems have increased the scale and application range of this paradigm: AgentSociety and its successor provide environments for large-scale computational social experiments \citep{piao2025agentsociety,piao2026agentsociety2}, and CitySim models urban schedules and collective city dynamics \citep{bougie2025citysim}.

These developments suggest several possible uses, provided that simulated behavior is not mistaken for validated human prediction. LLM simulations of established human-subject studies can reproduce some findings while also exhibiting systematic distortions \citep{aher2023simulate}. Studies that seek individual-level prediction require agents grounded in human data and evaluation against human responses \citep{park2024selfreports}. LLM-based simulations may support exploratory studies of communication, evacuation, public policy, urban activity, and collective response to external events. They may also help organizations prototype services, examine hypothetical customer or worker interactions, and identify scenarios that deserve later evaluation with real participants. Such simulations are therefore best treated as instruments for generating hypotheses, comparing mechanisms, and rehearsing possibilities, rather than as substitutes for empirical evidence.

Textual and two-dimensional environments capture communication and social organization efficiently, but they abstract away much of the spatial and perceptual structure of physical action. VLMs and vision-language-action (VLA) models provide a complementary direction by grounding decisions in images and language. PaLM-E demonstrated an embodied multimodal language model that integrates visual, textual, and continuous state inputs for embodied reasoning \citep{driess2023palme}. OpenVLA illustrates how visual and linguistic representations can be connected to actions in an open model for robot control \citep{kim2025openvla}. SimWorld exposes LLM and VLM agents to multimodal physical and social environments \citep{ren2025simworld}, CrowdVLA applies VLA agents to context-aware crowd simulation in semantically structured three-dimensional scenes \citep{hwang2026crowdvla}, and TravelAgent integrates generative agents into three-dimensional built environments for navigation and wayfinding \citep{noyman2024travelagent}. These embodied approaches may support exploratory work on navigation, facility use, human-robot interaction, crowd behavior, and digital twins.

The growing capability of such systems also creates an educational access problem. Large platforms, remote model services, specialized simulators, and substantial compute requirements can obscure the basic loop connecting perception, reasoning, communication, action, and observation. We present \agentfoundryname{}, a pair of minimal local simulation foundations comprising \agentfoundrytwoname{} for LLM-driven multi-agent social simulation and \agentfoundrythreename{} for VLM-driven embodied simulation. Our goal is not to provide another feature-complete simulator. Instead, we release two compact codebases that expose this loop directly and can serve as readable starting points. They use locally hosted models, include setup paths for major desktop operating systems, and retain deliberately simple mechanisms that users can replace. The intended contribution is an accessible foundation from which students, researchers, and practitioners can develop their own scenarios, agent designs, measurements, and interfaces.

\section{System Architecture}
\label{sec:method}

This section describes the architecture and implementation of \agentfoundrytwoname{} and \agentfoundrythreename{}.

\subsection{SD-AgentFoundry-2D}
\label{sec:method-llm-2d}

\begin{figure*}[!t]
\centering
\includegraphics[width=0.72\textwidth]{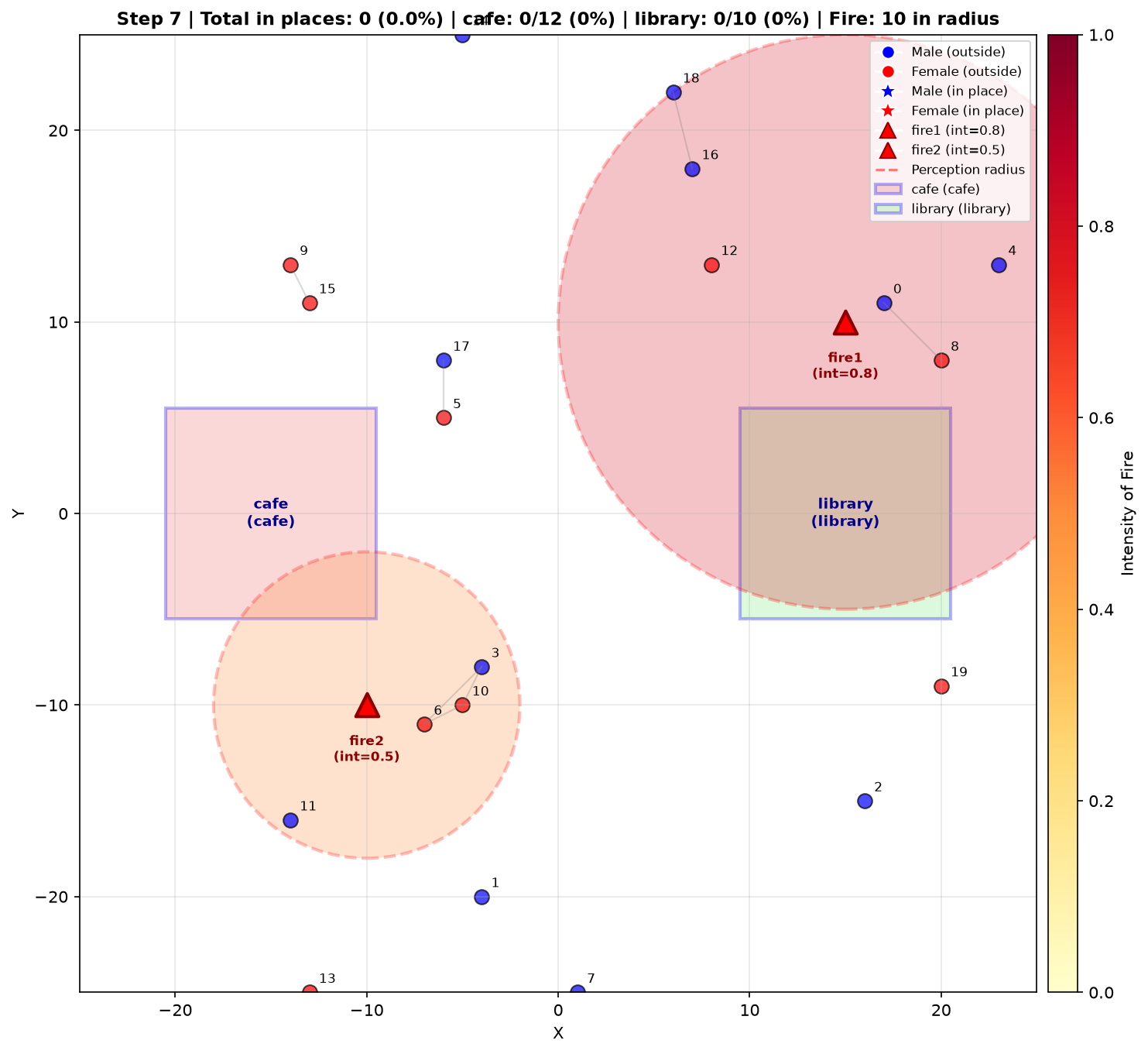}
\caption{Snapshot of an example \agentfoundrytwoplainname{} simulation. The two-dimensional grid contains 20 LLM-driven agents, two places (a cafe and a library), and two active fire events. Colored markers represent the agents, shaded disks show the fire-perception radii, and gray lines connect pairs eligible for local communication. Such configurable environments may support social-simulation studies and characterization of LLM-agent behavior.}
\label{fig:llm-2d-example}
\end{figure*}

\begin{figure*}[!t]
\centering
\includegraphics[width=0.78\textwidth]{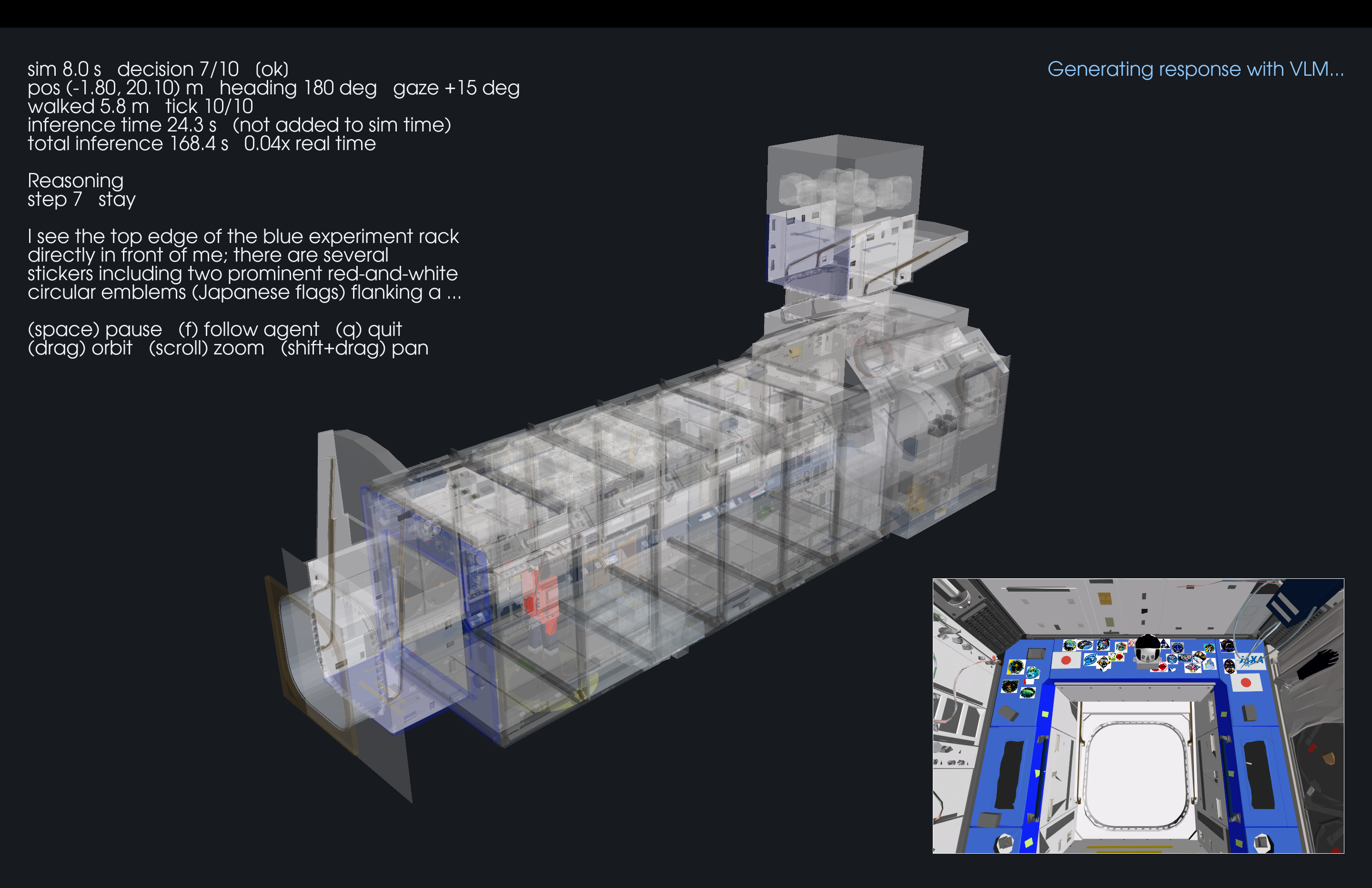}
\caption{Snapshot of an example \agentfoundrythreeplainname{} simulation in the Kibo digital twin. The Kibo module is loaded as the USD environment, where a single VLM-driven agent, shown as a red-clad human avatar, attempts the assigned task. The lower-right inset is a same-viewpoint re-rendering matched to the VLM camera pose and settings. Users can replace Kibo with their own USD scenes to run VLM-driven embodied simulations in custom three-dimensional spaces.}
\label{fig:vlm-3d-example}
\end{figure*}

\agentfoundrytwoname{} is a discrete two-dimensional multi-agent simulation implemented in Python (Figure~\ref{fig:llm-2d-example}). The environment is a bounded integer grid containing configurable places, each defined by a name, type, center, spatial extent, and capacity. The default scenario places a cafe and a library on opposite sides of the field. Agents begin outside the places at randomly generated positions and receive a simple persona field representing gender. Their available physical actions are to stay or move by one grid cell in one of four cardinal directions.

At every simulation step, each agent first identifies communicable neighbors. Communication requires both spatial proximity and a shared area: agents can communicate when they are outside all places together or when they occupy the same place, but not across place boundaries. Each agent then makes a message decision, the messages are delivered using the pre-movement neighborhood, each agent makes an action decision, and all selected movements are finally executed. Separating communication from movement gives every agent a consistent view of the current step.

\enlargethispage{\baselineskip}
Recent memories and received messages are retained in bounded histories and included in later prompts. Such bounded histories are a deliberate simplification of a much larger design space for agent memory, which ranges from parametric internal state to retrieval-augmented external stores \citep{huang2026agentmemory}.

The prompts provide numerical state rather than qualitative prescriptions. An agent inside a place receives its population, capacity, and occupancy rate, but it is not told that the place is comfortable or crowded. Configurable fire events begin at specified steps and have positions, intensities, and perception radii. Only agents inside a fire's radius receive its numerical properties and their distance from it; agents outside the radius can learn about the event only through messages. This design leaves interpretations such as avoidance, warning, coordination, or inaction to the LLM rather than encoding them as simulator rules. Related spatial LLM-agent studies have used capacity constraints and resource hazards to examine emergent, model-dependent behavior \citep{takata2025elfarol,masumori2025survival}.

The LLM is served locally through Ollama. One call determines a message and a second determines movement, memory, and reasoning, using JSON-shaped responses with conservative fallback parsing. Thus, each agent receives two distinct prompts per step: a communication prompt followed by an action-decision prompt. Detailed communication and action-decision prompts are shown in Appendix~\ref{sec:appendix-llm-prompts}. The current default uses a compact instruction-tuned Qwen model and disables model-side thinking so that the response budget is available for the requested JSON. The simulator writes memory and reasoning records to JSONL and creates a message log when communication occurs. A display-independent Matplotlib Agg renderer writes the initial state and selected post-step states as PNG files; it never opens a live window. The saved frames can be inspected directly, loaded in the separate browser viewer, or converted to video. Configuration is centralized in YAML, and setup scripts are supplied for macOS/Linux and Windows.

\subsection{SD-AgentFoundry-3D}
\label{sec:method-vlm-3d}

\agentfoundrythreename{} is a three-dimensional, single-agent simulation built around Universal Scene Description (USD) assets. The default demonstration loads a digital twin of the Japanese Experiment Module ``Kibo'' and gives the agent the task of locating the JAXA logo and approaching it (Figure~\ref{fig:vlm-3d-example}). The USD stage supplies geometry, scale, and coordinate information; the implementation uses \texttt{usd-core} for scene access and VTK/PyVista for rendering rather than requiring a heavyweight game engine.

\begin{table*}[!t]
\centering
\caption{Configuration and recorded artifacts for an example \agentfoundrytwoplainname{} run.}
\label{tab:llm-2d-settings}
\renewcommand{\arraystretch}{0.9}
\begin{tabularx}{\textwidth}{@{}lX@{}}
\toprule
Setting & Value \\
\midrule
LLM backend and model & Ollama; \texttt{qwen3:4b-instruct-2507-q4\_K\_M} \\
Inference settings & Temperature 0.2; thinking disabled; maximum 512 response tokens \\
Agents and inspected point & 20 agents; state recorded after step 7 \\
World & Integer grid with both axes ranging from $-25$ to $25$ \\
Places & Cafe at $(-15,0)$ and library at $(15,0)$, with capacities 12 and 10 \\
Fire events & \texttt{fire1}: step 3, intensity 0.8, radius 15; \texttt{fire2}: step 5, intensity 0.5, radius 8 \\
Step 7 state & Both fires active; 10 agents within at least one perception radius; 0 agents inside places \\
Recorded through step 7 & 8 PNG frames, 140 memory/reasoning records, and 88 per-recipient message-delivery records \\
\bottomrule
\end{tabularx}
\end{table*}

\begin{table*}[!t]
\centering
\caption{Configuration for an example \agentfoundrythreeplainname{} run.}
\label{tab:vlm-3d-settings}
\renewcommand{\arraystretch}{0.9}
\begin{tabularx}{\textwidth}{@{}lX@{}}
\toprule
Setting & Value \\
\midrule
VLM backend and model & Ollama; \texttt{qwen3.6:27b} \\
Scene & \path{assets/kibou/KIBOU.usd}, excluding the distant Earth backdrop \\
Task & Find JAXA logo and get close to it \\
Initial pose & Position $(4.0,20.1)$ m; yaw $180^{\circ}$; pitch $0^{\circ}$; eye height 1.6 m \\
Run length & 10 VLM decisions for the representative short run \\
Camera & $512\times384$ pixels; $70^{\circ}$ vertical field of view \\
Output & First-person frames, raw and parsed decisions, pose history, configuration snapshot, and replay frames \\
\bottomrule
\end{tabularx}
\end{table*}

At each decision point, the simulator renders the agent's first-person RGB image and sends that image and a dynamically assembled text prompt to a VLM. The text contains the task, current pose, and a bounded recent history. The detailed VLM prompt is shown in Appendix~\ref{sec:appendix-vlm-prompt}. The model returns unrestricted natural language rather than a required action schema. A deterministic interpreter scans the reply and resolves each movement category from its latest occurrence in the text, since models that reason aloud state their conclusion at the end. It converts supported distance units to meters and extracts translation bearing, body rotation, gaze pitch, or a stay command. Configured limits clamp excessive motion, and unreadable responses are recorded without inventing a replacement action. This interface is a VLM-driven, VLA-style perception--decision--action loop; it is not presented as a trained VLA policy.

The agent can translate, rotate its body, and change its gaze within configured bounds. A live view combines a semi-transparent overview of the scene, the agent and its trajectory, and an inset re-rendered from the same camera pose and settings as the first-person image supplied to the VLM. The default backend uses a local vision-capable model through Ollama, while local Transformers, OpenAI-compatible, and scripted backends provide replaceable interfaces. Inference wall-clock time is kept separate from simulation time so that model latency does not alter the simulated dynamics.

For reproducibility and auditing, every decision record preserves the raw model response, parsed command, parser notes, pre- and post-action poses, boundary effects, inference time, and the path of the input image when frame saving is enabled. The effective configuration is saved with the run. A separate replay path reconstructs overview frames from the log rather than treating the live window as a video recording. Cross-platform setup scripts are included.

\FloatBarrier

\section{Example Runs \& Beyond}
\label{sec:example-runs}

The following representative runs are intended as qualitative usage examples, not as controlled evaluations or evidence of predictive validity. Their tables document the configurations and recorded outputs. Exact repetition of the current 2D example additionally requires control of its randomly generated initial positions and personas, for which the present configuration does not expose a seed.

\subsection{Representative 2D LLM Run}
\label{sec:example-run-llm-2d}

This subsection presents an illustrative 2D simulation using \agentfoundrytwoname{}. Twenty LLM agents are placed in a grid world containing two accessible places, a cafe and a library (Figure~\ref{fig:llm-2d-example}). Agents can converse with nearby agents by exchanging natural-language messages. Two fires occur during the run, and each agent uses locally available information and received messages to decide how to move and what to communicate to others. The scenario therefore provides a compact setting in which to observe how LLM agents respond to a shared hazard, choose between places, and interact with one another. Holding the scenario fixed while replacing the underlying LLM can also reveal model-dependent differences in agent behavior.

The purpose of this example is to demonstrate the simulation workflow rather than to analyze its outcome in depth. Figure~\ref{fig:llm-2d-example} presents one snapshot of the run, showing the two fires and their perception radii, the spatial distribution of the 20 agents, and the pairs eligible for local communication. Table~\ref{tab:llm-2d-settings} summarizes the parameters and recorded outputs associated with this example. The snapshot confirms that the decision, communication, fire-perception, logging, and visualization components operate together, but it should not be interpreted as evidence of a causal behavioral pattern or as a validated model of human response because it represents only a single illustrative run rather than a dedicated behavioral study.

Related work has shown that two-dimensional or otherwise spatially explicit LLM-agent simulations can support the prototyping of populated social systems, the study of memory- and persona-conditioned behavior, and the observation of emergent interactions. Social Simulacra used simulated populations to prototype social-computing systems, while Generative Agents demonstrated socially interacting agents in a two-dimensional town, and Concordia provided a more general framework for language-mediated agent-based modeling \citep{park2022social,park2023generative,vezhnevets2023concordia}. Spatial extensions of the El Farol Bar problem have examined collective choice under capacity constraints \citep{takata2025elfarol}, while Sugarscape-style environments have revealed model-dependent behavior under resource scarcity and hazards \citep{masumori2025survival}. Larger platforms such as AgentSociety, AgentSociety 2, and CitySim further illustrate how generative agents may be used to explore collective behavior and urban dynamics \citep{piao2025agentsociety,piao2026agentsociety2,bougie2025citysim}. Accordingly, a simple two-dimensional world can serve both as a foundation for social-simulation experiments and as a controlled environment for studying the behavioral characteristics of LLM agents.

For such studies, researchers can, for example, vary persona distributions, initial layouts, place geometry and capacity, communication distance, memory length, inference settings, and the position, timing, intensity, and perception radius of hazards. More importantly, they can hold these conditions constant while changing the LLM and compare model-specific behavioral signatures, including movement preferences, action stability, sensitivity to hazards, communication frequency, warning propagation, clustering, and collective place choice. These measurements can help identify recurring tendencies or biases in how a particular LLM behaves when instantiated as an agent. They characterize behavior within the specified simulator and prompt design, however, rather than an intrinsic human-like personality or an empirically validated prediction of human populations. Rigorous comparison would additionally require controlled random seeds, repeated trials, and validation against appropriate human or observational data, as exemplified by work that evaluates human-grounded agents against participants' own responses \citep{park2024selfreports}.

\subsection{Representative 3D VLM Run}
\label{sec:example-run-vlm-3d}

This subsection presents an illustrative 3D simulation using \agentfoundrythreename{}. The Japanese Experiment Module ``Kibo'' is loaded as a USD scene, and a single VLM agent, represented by a human avatar in red clothing, is instructed to find the JAXA logo and approach it (Figure~\ref{fig:vlm-3d-example}). At each decision, the agent receives a first-person RGB image together with the task prompt and recent action history. The VLM describes the scene and proposes a movement in natural language, which the simulator interprets as a predefined translation, rotation, gaze change, or stay action. The scenario therefore provides a compact example of visual grounding and embodied decision making within a three-dimensional digital twin.

The purpose of this example is to demonstrate the simulation workflow rather than to analyze task performance in depth. Figure~\ref{fig:vlm-3d-example} presents one snapshot of the run: the overview shows the agent and its trajectory within the Kibo geometry, while the inset shows a same-viewpoint re-rendering matched to the VLM camera pose and settings. Table~\ref{tab:vlm-3d-settings} summarizes the parameters and recorded outputs associated with this example. The snapshot confirms that USD-scene loading, first-person rendering, VLM inference, action interpretation, movement, logging, and visualization operate together, but it should not be interpreted as a performance benchmark or as evidence of reliable task completion because it represents only a single illustrative run rather than a dedicated evaluation study.

Related work has shown that visually grounded agents in three-dimensional environments can support research on navigation, interaction, and behavior in physical and social spaces. Habitat established a configurable, photorealistic simulation platform for embodied-AI tasks such as navigation and instruction following \citep{savva2019habitat}. Vision-and-language navigation grounded natural-language navigation instructions in visual observations of previously unseen environments \citep{anderson2018vln}. More recent examples include SimWorld, which provides multimodal environments for autonomous agents in open-ended physical and social scenarios \citep{ren2025simworld}; TravelAgent, which studies navigation and wayfinding by generative agents in built environments \citep{noyman2024travelagent}; and CrowdVLA, which addresses context-aware agent navigation and continuous locomotion in semantically structured crowd simulations \citep{hwang2026crowdvla}. EmbRACE-3K further provides a benchmark for first-person, multi-step embodied reasoning and reports substantial limitations in the zero-shot task success of contemporary VLMs \citep{lin2025embrace3k}.

These studies illustrate how three-dimensional simulation can connect language-based reasoning with spatial perception and action. Accordingly, a configurable USD world can serve both as a foundation for digital-twin experiments and as a controlled environment for studying the embodied behavioral characteristics of vision-language agents.

For such studies, researchers can, for example, replace the Kibo asset with another USD scene and vary the task instruction, initial pose, camera field of view, movement and gaze limits, recent-history length, model backend, and response interpreter. Facilities, streets, workplaces, or other digital twins could thereby be used to study visual grounding, object search, wayfinding, viewpoint dependence, and sensitivity to language or scene layout. More importantly, researchers can hold the scene, task, and action constraints constant while changing the VLM and compare model-specific behavioral signatures, including object-recognition errors, exploration strategies, movement efficiency, action consistency, recovery from incorrect decisions, and sensitivity to viewpoint. Perspective-taking is also an explicit dimension of comprehensive VLM spatial-reasoning benchmarks \citep{jia2026omnispatial}. These measurements characterize behavior within the specified simulator, prompt, and action interface rather than general visual intelligence or real-world embodied competence. Rigorous comparison would additionally require controlled initial conditions, repeated trials, task-specific success criteria, and validation in appropriate physical or observational settings.

The present VLM-driven loop should not be confused with a trained VLA policy. Here, a VLM produces natural-language text, and a deterministic interpreter maps recognized phrases onto a predefined set of translations, rotations, gaze changes, and stay actions. In contrast, VLA models such as RT-2 and OpenVLA are trained with robot trajectories or demonstrations so that visual and linguistic inputs are connected directly to embodiment-specific actions and transferable visuomotor skills \citep{brohan2023rt2,kim2025openvla}. CrowdVLA similarly introduces learned motion skills to bridge symbolic decisions and continuous locomotion in crowd simulation \citep{hwang2026crowdvla}.

Consequently, \agentfoundrythreename{} is suitable for interpretable studies of perception, high-level decision making, and bounded simulated navigation, but it cannot by itself learn a new motor skill, generate joint- or end-effector-level control, handle contact-rich manipulation, or close a high-frequency sensorimotor loop. Those capabilities require a VLA or another trained low-level control policy, an embodiment-specific interface, and separate physical validation. Generative Pretrained Controllers exemplify a distinct learned low-level policy that maps reusable motion representations to physics-based character control \citep{shi2026gpc}.

\FloatBarrier
\section{Summary}
\label{sec:summary}

LLM- and VLM-driven agents make it possible to explore social interaction and embodied decision making with mechanisms that are difficult to express as fixed behavioral rules. Such simulations may inform hypothesis generation for social challenges, urban and facility planning, emergency communication, service prototyping, and other business scenarios, but their outputs require empirical validation before they can support real-world claims. The two \agentfoundryname{} foundations expose complementary levels of abstraction: \agentfoundrytwoname{} emphasizes communication and collective behavior in a transparent two-dimensional world, whereas \agentfoundrythreename{} emphasizes visual grounding and physical action in a three-dimensional digital twin.

Both systems are intentionally basic, locally operated, and designed for use on macOS, Windows, and Linux. Their purpose is educational accessibility and extensibility rather than completeness. By publishing readable implementations of the perception--reasoning--action loop, configurable scenarios, visualizations, and auditable logs, we hope to lower the barrier to studying these techniques and to encourage users to add their own environments, agent architectures, models, measurements, and application-specific safeguards.

\section*{Acknowledgment}

R.H. acknowledges financial support from JSPS KAKENHI Grant Numbers 26K00756, 23KK0253, 22K14091, 21H04512, 21H04514, and 20KK0080. ChatGPT was used to assist with English-language proofreading of the manuscript.

\enlargethispage{2\baselineskip}
\printbibliography[title={References}]
\flushcolsend

\clearpage
\onecolumn
\appendix

\section{Prompts Used in the Simulations}
\label{sec:prompt-appendix}

This appendix documents the text supplied to the local models. Angle-bracketed expressions (e.g., \texttt{<xxx>}) in the templates denote values inserted at runtime. Blocks explicitly marked as conditional are omitted when the corresponding information is unavailable. Line breaks inside the framed listings are part of the prompts, whereas automatic visual wrapping does not insert additional characters.

\subsection{SD-AgentFoundry-2D LLM Prompts}
\label{sec:appendix-llm-prompts}

Each agent makes two sequential LLM calls during a simulation step. The first, a communication prompt (Section~\ref{sec:appendix-communication-prompt}), determines what the agent says to locally communicable agents and deliberately excludes their coordinates. The selected messages are delivered before movement. The second, an action-decision prompt (Section~\ref{sec:appendix-action-prompt}), includes spatial coordinates and the exchanged messages and requests a movement or stay action, memory, and reasoning. These two sections reproduce the fixed text in the implementation and identify every dynamic or conditional field. Section~\ref{sec:appendix-example-action-prompt} then provides a reconstructed example showing how representative runtime values populate the action-decision template.

\subsubsection{Communication prompt template}
\label{sec:appendix-communication-prompt}

\begin{lstlisting}[style=promptstyle]
You are Agent <agent-id> (<gender>) in a 2D world with multiple places (<comma-separated unique place types>).

=== YOUR CURRENT STATE ===
Gender: <gender>
In place: <Yes or No>
<Current place: place-name; omitted when outside>

<CONDITIONAL WHEN INSIDE A PLACE>
You are currently in the <place-type> (<place-name>).
  Number of agents here: <agents-in-place>
  Capacity: <capacity>
  Occupancy rate: <occupancy-rate rounded to two decimals>

<CONDITIONAL WHEN ONE OR MORE FIRES ARE PERCEIVED>
=== FIRE EVENT ===
<REPEATED FOR EACH PERCEIVED FIRE>
Fire "<fire-name>":
  Position: (<fire-x>, <fire-y>)
  Intensity: <intensity> (scale: 0.0 to 1.0)
  Radius: <radius>
  Your distance: <distance rounded to two decimals>

=== NEARBY AGENTS (you can communicate with these agents) ===
<Agent id (gender) is in place-name (place-type), Agent id (gender) is outside the places, or No nearby agents.>

=== PREVIOUS MEMORY ===
<- one line per retained memory, or No previous experiences.>

=== MESSAGES FROM OTHERS ===
<from Agent id: message, or No messages received.>

=== YOUR TASK ===
Decide what message you want to send to nearby agents. You can share your observations, experiences, or thoughts about the places and situation.

=== RESPOND IN JSON ===
{
    "message": "message to nearby agents (max 200 words, optional if you don't want to send a message)",
    "reasoning": "brief explanation of why you want to send this message"
}

Step: <step>
\end{lstlisting}

\subsubsection{Action-decision prompt template}
\label{sec:appendix-action-prompt}

\begin{lstlisting}[style=promptstyle]
You are Agent <agent-id> (<gender>) in a 2D world with multiple places (<comma-separated unique place types>).

=== YOUR CURRENT STATE ===
Gender: <gender>
Position: (<x>, <y>)
In place: <Yes or No>
<Current place: place-name; omitted when outside>

<CONDITIONAL WHEN INSIDE A PLACE>
You are currently in the <place-type> (<place-name>).
  Number of agents here: <agents-in-place>
  Capacity: <capacity>
  Occupancy rate: <occupancy-rate rounded to two decimals>

<CONDITIONAL WHEN ONE OR MORE FIRES ARE PERCEIVED>
=== FIRE EVENT ===
<REPEATED FOR EACH PERCEIVED FIRE>
Fire "<fire-name>":
  Position: (<fire-x>, <fire-y>)
  Intensity: <intensity> (scale: 0.0 to 1.0)
  Radius: <radius>
  Your distance: <distance rounded to two decimals>

=== PLACE LOCATIONS ===
<place-name> (<place-type>): center at (<center-x>, <center-y>), covers X from <minimum-x> to <maximum-x>, Y from <minimum-y> to <maximum-y>
<one line for every configured place>

=== NEARBY AGENTS ===
<Agent id (gender) is at (x, y) and is in place-name (place-type), Agent id (gender) is at (x, y) and is outside the places, or No nearby agents.>

=== PREVIOUS MEMORY ===
<- one line per retained memory, or No previous experiences.>

=== MESSAGES FROM OTHERS ===
<from Agent id: message, or No messages received.>

<CONDITIONAL WHEN A MESSAGE WAS CHOSEN>
=== MESSAGE YOU DECIDED TO SEND ===
<message selected by the communication call>
=== AVAILABLE ACTIONS ===
- "stay": remain at current position
- "move" with direction: "up" (Y+1), "down" (Y-1), "left" (X-1), "right" (X+1)

Field boundaries: X and Y from -<half-space-size> to +<half-space-size>

=== RESPOND IN JSON ===
{
    "action": "move" or "stay",
    "direction": "up", "down", "left", or "right" (only if action is "move"),
    "memory": "what you want to remember for the next step (your thoughts, observations, intentions)",
    "reasoning": "brief explanation of your decision"
}

Step: <step>
\end{lstlisting}

\subsubsection{Example action prompt}
\label{sec:appendix-example-action-prompt}

The 2D implementation does not retain the exact prompts sent to Ollama. The following is therefore not a historical record or an observed result. It is a representative prompt generated with the inspected code and current cafe/library place definitions at step~7, after \texttt{fire1} has activated. Female Agent~3 is at $(14,2)$ in the library with two other occupants, one retained memory, one received message, nearby Agent~1, and perceived fire event \texttt{fire1} at a distance of 8.06 grid units. The order in which distinct place types appear in the opening sentence can vary because the implementation derives that list through set iteration.

\begin{lstlisting}[style=promptstyle]
You are Agent 3 (female) in a 2D world with multiple places (cafe, library).

=== YOUR CURRENT STATE ===
Gender: female
Position: (14, 2)
In place: Yes
Current place: library

You are currently in the library (library).
  Number of agents here: 3
  Capacity: 10
  Occupancy rate: 0.30

=== FIRE EVENT ===
Fire "fire1":
  Position: (15, 10)
  Intensity: 0.8 (scale: 0.0 to 1.0)
  Radius: 15
  Your distance: 8.06

=== PLACE LOCATIONS ===
cafe (cafe): center at (-15, 0), covers X from -20 to -10, Y from -5 to 5
library (library): center at (15, 0), covers X from 10 to 20, Y from -5 to 5

=== NEARBY AGENTS ===
Agent 1 (male) is at (16, 1) and is in library (library)

=== PREVIOUS MEMORY ===
- I entered the library to inspect its occupancy.

=== MESSAGES FROM OTHERS ===
from Agent 1: I can also see fire1 from inside the library.

=== MESSAGE YOU DECIDED TO SEND ===
I can detect fire1 near the library; please stay alert.
=== AVAILABLE ACTIONS ===
- "stay": remain at current position
- "move" with direction: "up" (Y+1), "down" (Y-1), "left" (X-1), "right" (X+1)

Field boundaries: X and Y from -25 to +25

=== RESPOND IN JSON ===
{
    "action": "move" or "stay",
    "direction": "up", "down", "left", or "right" (only if action is "move"),
    "memory": "what you want to remember for the next step (your thoughts, observations, intentions)",
    "reasoning": "brief explanation of your decision"
}

Step: 7
\end{lstlisting}

\subsection{SD-AgentFoundry-3D VLM Prompt}
\label{sec:appendix-vlm-prompt}

The VLM receives a dynamically assembled text prompt together with the current first-person RGB image. Section~\ref{sec:appendix-vlm-template} presents the text-prompt template. Its recent-history block is limited by the configured memory size; each entry contains the earlier raw reply flattened to one line and truncated to at most 160 characters, the interpreter's outcome, and the resulting pose. The final nudge is included only when the latest three retained commands are all idle. Section~\ref{sec:appendix-vlm-example} provides an example assembled for decision step~0 of the current Kibo configuration, when no action history is yet available.

\subsubsection{Dynamic text-prompt template}
\label{sec:appendix-vlm-template}

\begin{lstlisting}[style=promptstyle]
<task prompt from config.yaml>

Movement conventions: distances are in metres unless you name another unit; angles are in degrees. Turning left is counter-clockwise. You can look up or down, but not straight up or straight down.

Your current pose: position (<x>, <y>) m, heading <yaw> deg, gaze <signed-pitch> deg.

<CONDITIONAL WHEN RECENT HISTORY EXISTS>
What you did recently:
- step <decision-index>: you said "<flattened reply>" -> <interpreted outcome>. You were then at <pose summary>.
<one line per retained history entry>

<CONDITIONAL AFTER THREE IDLE DECISIONS>
You have not moved for several steps. Try something different.

What is your next single movement?
\end{lstlisting}

\subsubsection{Example text prompt}
\label{sec:appendix-vlm-example}

The following is the text assembled for decision step~0 when the current task prompt in \texttt{config.yaml} is combined with the logged initial pose $(x,y,z)=(4.0,20.1,-0.2358)$, yaw $180^{\circ}$, and pitch $0^{\circ}$. The prompt exposes only the two-dimensional foot position, heading, and gaze. Because no earlier decision exists at step~0, there is no history block. The VLM receives this text together with the corresponding first-person RGB image.

\begin{lstlisting}[style=promptstyle]
You are an agent standing inside the Japanese Experiment Module "Kibo" of the ISS.
The image is your own first-person view.
Task: Find JAXA logo and get close to it.
Briefly say what you see and why you are moving, then state your next single
movement in one sentence, LAST.
Examples of the movement sentence: "move forward 2 m", "turn right 30 degrees",
"look up 30 degrees", "stay".

Movement conventions: distances are in metres unless you name another unit; angles are in degrees. Turning left is counter-clockwise. You can look up or down, but not straight up or straight down.

Your current pose: position (4.00, 20.10) m, heading 180 deg, gaze +0 deg.

What is your next single movement?
\end{lstlisting}

\end{document}